\documentclass[twocolumn,aps,pra,10pt]{revtex4-2}
\usepackage{bm}
\usepackage{amsfonts}
\usepackage{amssymb}
\usepackage{amsmath}
\usepackage{graphicx}
\begin{document}

\title{The Zeldovich number: A universal dimensionless measure\\for the electromagnetic field}
\author{Iwo Bialynicki-Birula}\email{birula@cft.edu.pl}
\affiliation{Center for Theoretical Physics, Polish Academy of Sciences\\
Aleja Lotnik\'ow 32/46, 02-668 Warsaw, Poland}
\author{Zofia Bialynicka-Birula}
\affiliation{Institute of Physics, Polish Academy of Sciences\\
Aleja Lotnik\'ow 32/46, 02-668 Warsaw, Poland}
\date{\today}

\begin{abstract}
In this work we extend the Zeldovich formula, which was originally derived for the free electromagnetic field and was interpreted as the number of photons. We show that our extended formula gives a universal dimensionless measure of the overall strength of electromagnetic fields: free fields and fields produced by various sources, in classical and in quantum theory. In particular, we find that this number (the Zeldovich number) for macroscopic systems is huge, of the order of $10^{20}$. For the hydrogen atom in the ground state it is equal to 0.025 and for the xenon atom it is around 50.
\end{abstract}

\maketitle

\section{Introduction}

The formula measuring the number of photons was derived by Zeldovich for the free electromagnetic field. In the original derivation \cite{zeld} of this formula, Zeldovich assumed that photons can be identified with monochromatic oscillations of the electromagnetic field. However, we will show that this formula is universal; it can be used without any restrictions for all electromagnetic fields. In the quantum theory of the electromagnetic field the Zeldovich formula plays multiple roles. It can be used to measure the probability of various field configurations and to define the norm of quantum photon states.

In this work we extend the use of the Zeldovich formula from free fields to field configurations generated by various sources. In particular, we will study the problem of the photons attached to the hydrogen atom which has been treated in nonrelativistic case by Francesco Persico and his collaborators \cite{pcp0,pcp,cppp,cpps}. The extension to the relativistic theory was treated in \cite{tr}. All these attemps have not given a definite answer to the question: How many photons are attached to an atom? We obtain an answer to this question employing the notion of the Zeldovich number.

Since in the general case the name {\em photon number} is not always justified, we shall use instead the term {\em Zeldovich number} and we denote it by $\mathcal{N}_Z$. The number $\mathcal{N}_Z$ is a {\em dimensionless} quantity which measures the overall strength of the electromagnetic field. This number is a useful characterization of the field and its sources.

In Section II we extend the derivation of the formula for $\mathcal{N}_Z$ to a general electromagnetic field. In Section III we calculate $\mathcal{N}_Z$ in classical electrodynamics. In Sections IV and V we extend the calculations to atomic physics, nonrelativistic and relativistic. This is done by associating the wave functions of electrons with electromagnetic fields.\\

\section{The Zeldovich number}

The starting point of our calculations is the formula for the total energy of the electromagnetic field,
\begin{align}\label{en}
\mathcal{E}(t)=\frac{1}{2}
\int\!d^3r\left[
{\bm E}({\bm r},t)\cdot{\bm D}({\bm r},t)+{\bm B}({\bm r},t)\cdot\!{\bm H}({\bm r},t)\right].
\end{align}
We do not assume, as has been done by Zeldovich, that the field vectors satisfy the free Maxwell equations. It is sufficient that the field vectors are sufficiently regular to have Fourier transforms.

It is convenient to use the field vectors ${\bm{\mathcal D}}$ and ${\bm{\mathcal H}}$ measured in purely geometrical units, i.e. in $1/{\rm meter^{2}}$. They are related to the standard physical vectors ${\bm{D}}$ and ${\bm{H}}$ by the formulas,
\begin{align}\label{units}
{\bm{\mathcal D}}=\frac{\bm D}{e},\quad{\bm{\mathcal H}}=\frac{\bm H}{e\,c}.
\end{align}
The expression for the field energy expressed in terms of the new vectors is,
\begin{align}\label{enn}
\mathcal{E}(t)=\frac{e^2}{2\epsilon_0}\!
\int\!\!d^3r\left[
\bm{\mathcal D}({\bm r},t)\!\cdot\!\bm{\mathcal D}({\bm r},t)+\bm{\mathcal H}({\bm r},t)\!\cdot\!\bm{\mathcal H}({\bm r},t)\right].
\end{align}
Next, we rewrite this formula in terms of Fourier transforms,
\begin{align}\label{enf}
\mathcal{E}(t)=\frac{e^2}{2\epsilon_0}\!
\int\!\!d^3k\left[
\bm{\tilde{\mathcal D}^*}({\bm k},t)\!\cdot\!\bm{\tilde{\mathcal D}}({\bm k},t)+\bm{\tilde{\mathcal H}^*}({\bm k},t)\!\cdot\!\bm{\tilde{\mathcal H}}({\bm k},t)\right].
\end{align}
The original formula for $\mathcal{N}_Z$ will be obtained in two steps. In the first step we divide the integrand in (\ref{enf}) by $\hbar c k$ to obtain a dimensionless quantity,
\begin{widetext}
\begin{align}\label{nf}
\mathcal{N}_Z\left[\bm{\mathcal D},\bm{\mathcal H}\right]=2\pi\alpha\!
\int\!\!\frac{d^3k}{k}\left[
\bm{\tilde{\mathcal D}^*}({\bm k},t)\!\cdot\!\bm{\tilde{\mathcal D}}({\bm k},t)+\bm{\tilde{\mathcal H}^*}({\bm k},t)\!\cdot\!\bm{\tilde{\mathcal H}}({\bm k},t)\right],
\end{align}
\end{widetext}
where $\alpha$ is the fine structure constant. This is the representation in terms of Fourier transforms. In the next step, to obtain the original form of $\mathcal{N}_Z$ we convert (\ref{nf}) back to the $\bm r$ space with the use of relations,
\begin{align}\label{ft}
\bm{\tilde{\mathcal D}}({\bm k},t)=\int\!\frac{d^3r}{(2\pi)^{3/2}}e^{-i{\bm k\cdot{\bm r}}}\,\bm{\mathcal D}({\bm r},t),\\
\bm{\tilde{\mathcal H}}({\bm k},t)=\int\!\frac{d^3r}{(2\pi)^{3/2}}e^{-i{\bm k\cdot{\bm r}}}\,\bm{\mathcal H}({\bm r},t),
\end{align}
and the formula,
\begin{align}\label{form}
\int\!\frac{d^3k}{k}e^{i{\bm k}\cdot({\bm r}-{\bm r}')}=\frac{1}{2\pi^2\vert{\bm r}-{\bm r}'\vert^2}.
\end{align}
The resulting expression has the form obtained by Zeldovich  (apart from our different scaling of the electromagnetic field),
\begin{widetext}
\begin{align}\label{avn}
\mathcal{N}_Z\left[\bm{\mathcal D},\bm{\mathcal H}\right]=\frac{\alpha}{\pi}
\int\!d^3r\!\!\int\!d^3r'\!\left(
\bm{\mathcal D}({\bm r},t)\frac{1}{\vert{\bm r}-{\bm r}'\vert^2}\!\cdot\!\bm{\mathcal D}({\bm r}',t)+\bm{\mathcal H}({\bm r},t)\frac{1}{\vert{\bm r}-{\bm r}'\vert^2}\!\cdot\!\bm{\mathcal H}({\bm r}',t)\right).
\end{align}
\end{widetext}
The derivation of this expression by Zeldovich was based on the expansion of the solutions of Maxwell equations in free space into monochromatic propagating waves. We arrived at the same formula without making any assumptions concerning the dynamics of the electromagnetic field. Of course, the interpretation of $\mathcal{N}_Z$ in this general case as the photon number is highly problematic, but $\mathcal{N}_Z$ is very well defined.

In the case of free fields, $\mathcal{N}_Z$ is the total number of photons and it has some remarkable properties. It is a constant of motion and despite its nonrelativistic appearance it is invariant not only under all Lorentz transformations, but also under the conformal transformations \cite{gross}. Since conformal transformations include the transformations to accelerated frames of reference, the invariance of $\mathcal{N}_Z$ may help to better understand the Unruh effect \cite{rmp}. As a mathematical object, $\mathcal{N}_Z$ plays the role of a norm for the photon wave function \cite{gross,pwf,bb1}. The scalar product obtained from this norm by polarization identity serves as a perfect measure of fidelity for photon states \cite{bb}. It also appears as the exponent in the Wigner functional of the electromagnetic field \cite{ibb1,ibb2,ibb3}. In this way it determines the relative probabilities of various field configurations. Due to its connection to the Wigner functional, $\mathcal{N}_Z$ can be generalized to characterize also thermal states of electromagnetic fields \cite{ibb2,ibb3}.

\section{The Zeldovich number for macroscopic fields}

In this Section we calculate $\mathcal{N}_Z$ for field configurations created by the following classical sources: two oppositely charged metallic spheres and the current flowing in a circular loop.

In the first case the charged density is,
\begin{align}\label{den}
\rho({\bm r})=\frac{Q}{4\pi a^2}\left[\delta(a-|\bm{r+d/2}|)
-\delta(a-|\bm{r-d/2}|)\right],
\end{align}
where $a$ is the sphere radius, $|\bm d|$ is the distance between the spheres, and $Q$ is the charge. We assumed that the charge is distributed uniformly on the surface. We will choose the $z$ direction along the vector $\bm d$ and use spherical coordinates. To calculate $\mathcal{N}_Z$ we will need the Fourier transform $\tilde{\rho}(\bm k)$,
\begin{align}\label{denf}
\tilde{\rho}(\bm k)&=
\int\!\frac{d^3r}{(2\pi)^{3/2}}\rho({\bm r})e^{-i{\bm k\cdot{\bm r}}}\nonumber\\&=\frac{2iQ}{(2\pi)^{3/2}}\sin(d k/2\cos\theta)\frac{\sin(ak)}{ak},
\end{align}
where we used the property that a shift by $\pm \bm d/2$ in the position space results in the multiplication by $e^{\pm i{\bm k\cdot{\bm d}/2}}$ in the Fourier space. For electrostatic fields, the Fourier transform of the displacement vector ${\mathcal D}$ is,
\begin{align}\label{d}
e\bm{\tilde{\mathcal D}}({\bm k})=-i\frac{{\bm k}\tilde{\rho}(\bm k)}{k^2}.
\end{align}
The substitution of this expression into (\ref{nf}) gives,
\begin{align}\label{eg1}
\!\mathcal{N}_Z=2\pi\alpha\left(\frac{Q}{e}\right)^2\!\!\!
\int_0^\infty\!\!\frac{dk}{k}\int_0^\pi\!\!\!d\theta\sin\theta\!\! \int_0^{2\pi}\!\!\!\!d\phi\,
\tilde{\rho}^*({\bm k})\tilde{\rho}({\bm k}).
\end{align}
These integrals can be analytically calculated and the final result, as was to be expected, depends only on the dimensionless ratio $b=d/a$,
\begin{widetext}
\begin{align}\label{b}
\mathcal{N}_Z=\frac{\alpha}{12\pi b}\left(\frac{Q}{e}\right)^2\left[(b+2)^3\ln(b+2)+(b-2)^3\ln(|b-2|)-2b(4+12\ln2+b^2\ln b\right].
\end{align}
\end{widetext}
When $b$ is very large, we obtain,
\begin{align}\label{lb}
\mathcal{N}_Z\approx\frac{2\alpha}{\pi}\left(\frac{Q}{e}\right)^2\ln b.
\end{align}
The unbounded logarithmic growth of $\mathcal{N}_Z$ is the mani\-festation of the infrared catastrophe, well known in quantum electrodynamics. In the case of a large separation the Coulomb fields of each sphere is practically not shielded and $\mathcal{N}_Z$ for an unshielded charge is infinite.

In the second example the current flowing in a closed loop has the following density in cylindrical coordinates \cite{jack},
\begin{align}\label{curc}
{\bm j}(\rho,z,\phi)
=I\delta(a-\rho)\delta(z)\{-\sin\phi,\cos\phi,0\}.
\end{align}
To calculate $\mathcal{N}_Z$, we will need the Fourier transform of the current,
\begin{align}\label{curcf}
&\tilde{\bm j}(\bm k)
=\int\frac{d^3r}{(2\pi)^{3/2}}e^{-i{\bm k\cdot{\bm r}}}{\bm j}(\rho,z,\phi)\nonumber\\
&=\frac{aI}{(2\pi)^{3/2}}\int_0^{2\pi}\!\!d\phi\,e^{-ia(k_x\cos\phi+ky\sin\phi)}
\{-\sin\phi,\cos\phi,0\}\nonumber\\
&=\frac{iaI}{\sqrt{2\pi}}\frac{J_1\left(ak_\perp\right)}{k_\perp}
\{k_y,-k_x,0\},
\end{align}
where $k_\perp=\sqrt{k_x^2+k_y^2}$ and $J_1$ is the Bessel function. For static fields, the Fourier transform of the magnetic field vector is,
\begin{align}\label{h}
ec\bm{\tilde{\mathcal H}}({\bm k})
=-i\frac{{\bm k}\times\tilde{\bm j}(\bm k)}{k^2}.
\end{align}
The substitution of this expression into (\ref{nf}) gives,
\begin{align}\label{eg2}
\mathcal{N}_Z&=\alpha\left(\frac{aI}{ec}\right)^2\!\!
\int_0^\infty\!\!\!dk_\perp\!\! \int_{-\infty}^\infty\!\!\!\!dk_z \int_0^{2\pi}\!\!\!\!d\phi
\frac{k_\perp J_1\left(ak_\perp\right)^2}
{(k_\perp^2+k_z^2)^{3/2}}\nonumber\\
&=4\pi\alpha\left(\frac{aI}{ec}\right)^2.
\end{align}
Note that the increase of the ring radius and the proportional decrease of the current leaves the Zeldovich number unchanged.

For macroscopic systems $\mathcal{N}_Z$ is huge. In the electrostatic case even for tiny charges of one microcoulomb on both spheres and for $b=10$ we obtain $\mathcal{N}_Z$ equal to $1.6\times10^{20}$. In turn, for the current of one ampere flowing in a loop with the radius of one meter we obtain $\mathcal{N}_Z$ equal to $4\times 10^{19}$. These very large values are due to the mismatch between the
value of elementary charge which appears in the definition of $\mathcal{N}_Z$ and these values in macroscopic fields.\\

\section{The Zeldovich number\\for the hydrogen atom}

In our calculations of $\mathcal{N}_Z$ associated with the hydrogen atom we will use the ground state electron wave function satisfying the Dirac equation. The wave functions satisfying the nonrelativistic Schr\"odinger equation would be simpler, but the relativistic treatment allows for a uniform treatment of the electric and magnetic fields.

The ground state is doubly degenerate (disregarding tiny corrections due to the hyperfine interactions). The two states differ in the sign of the projection of the total angular momentum on a chosen direction. We choose, as is customary, the $z$ direction. The state with the positive sign has the following normalized Dirac wave function \cite{bd},
\begin{align}\label{wf}
\psi(x,y,z,t)=\sqrt{\frac{\gamma+1}{8\pi\Gamma(2\gamma+1)}}
\left(\frac{2\alpha}{\lambdabar}\right)^{3/2}e^{-iEt}\nonumber\\
\times\left(\frac{2\alpha r}{\lambdabar}\right)^{\gamma-1}\!\!\!\!\!e^{-\alpha r/\lambdabar}
\left\{1,0,\frac{i\alpha}{\gamma+1}\frac{z}{r},\frac{i\alpha}{\gamma+1}
\frac{x+iy}{r}\right\},
\end{align}
where $\gamma=\sqrt{1-\alpha^2}$ and $\lambdabar=\hbar/mc$ is the reduced electron Compton wave length.

The probability density $\rho_e=\psi^*\!\!\,\psi$ and the probability current density ${\bm j}_e=\psi^*{\bm\alpha}\,\psi$ for the electron are,
\begin{align}
\rho_e(r)&=\frac{e^{-2r/b}(2r/b)^{2\gamma+1}}
{4\pi r^3\Gamma(2\gamma+1)}
,\label{rho}\\
{\bm j}_e({\bm r})&=\{-y,x,0\}\,\frac{\alpha\rho_e(r)}{r}\label{cur},
\end{align}
where  $b=\lambdabar/\alpha$ is the Bohr radius. These sources produce the electromagnetic field $\bm{\mathcal D}$ and $\bm{\mathcal H}$ generated by the electron. Since we are using the rescaled electric and magnetic field, the sources (\ref{rho}) and (\ref{cur}) must also be rescaled. The substitution of the electronic $\bm{\mathcal D}$ into (\ref{nf}) produces the infrared divergence. This divergence has a clear physical interpretation analogous to the one encountered for two charged spheres. The infinite result is simply due to the unshielded electron charge. Atoms are neutral and the introduction of the compensating charge of the nucleus will remove the infrared divergence.

The exact formula for the charge distribution in the nucleus is not important. The tiny size of the nucleus as compared to the size of the electronic cloud makes the difference between various models of the nucleus negligible. In our calculations of the electromagnetic field associated with the hydrogen atom we have assumed that the proton charge and the proton magnetic moment are distributed uniformly within a sphere with a sharp boundary (cf. \cite{march}) with the radius $a$ taken from experiment. Thus, the proton charge density and the current density will be described by the Heaviside step function,
\begin{align}
\rho_p(r)&=\frac{3}{4\pi a^3}\Theta(a-r),\label{rhop}\\
{\bm j}_p({\bm r})&=\frac{\{y,-x,0\}}{r}\frac{3\mu}{\pi a^4 }\Theta(a-r),\label{curp}
\end{align}
where $\mu$ in our geometrical units measures the strength of the proton magnetic moment $\mu_p$,
\begin{eqnarray}
\mu=\mu_p/(e c)=5.8\times10^{-16}\rm{m}.
\end{eqnarray}
These densities satisfy the conditions that the electric and magnetic fields outside the proton are correct,
\begin{align}\label{out}
{\bm{\mathcal D}_{\rm out}}({\bm r})&=\frac{\bm{r}}{4\pi r^3},\\
{\bm{\mathcal H}_{\rm out}}({\bm r})&=\frac{\mu}{4\pi r^3}\left(\frac{3{\bm r}({\bm r}\cdot{\bm n}_z)}{r^2}-{\bm n}_z\right),
\end{align}
where ${\bm n}_z$ is the unit vector along the $z$ axis.
We assumed that the proton is at rest. Hence the only contribution to its current comes form the proton magnetic moment $\mu$.

There are three scale parameters in our problem separated by a few orders of magnitude: the proton radius $a=8.5\times 10^{-16}$\,m,\, the Compton wave length of the electron $\lambdabar=3.86\times 10^{-13}$\,m,\, and the atomic scale parameter, the Bohr radius $b=5.29\times 10^{-11}$\,m.

There is a strong dependence of the Zeldovich number on the proton radius, which is fixed by the experiment. Owing to the smallness of the proton radius as compared to $\lambdabar$ and $b$ the eventual changes of the charge distribution inside the proton have almost no influence.

The field vectors ${\bm{\mathcal D}}({\bm r})$ and ${\bm{\mathcal H}}({\bm r})$ are the solutions of the Maxwell equations,
\begin{align}\label{max}
\bm{\nabla}\cdot{\bm{\mathcal D}}({\bm r})=\rho(r),\quad
\bm{\nabla}\times{\bm{\mathcal H}}({\bm r})={\bm j}({\bm r}),
\end{align}
where
\begin{align}\label{src}
\rho(r)=\rho_p(r)-\rho_e(r),\quad{\bm j}({\bm r})={\bm j}_p({\bm r})-{\bm j}_e({\bm r}).
\end{align}
Note that the elementary charge $e$ does not appear explicitly in these equations since we are using the field vectors in the geometrical units defined in (\ref{units}). The terms with time derivatives do not appear in (\ref{max}) because the sources do not depend on time.

The solutions of the Maxwell equations are most easily obtained with the use of two scalar functions $\phi(r)$ and $\mathfrak{a}(r)$,
\begin{align}
{\bm{\mathcal D}}({\bm r})&=-\bm{\nabla}\phi(r),\label{mpot1}\\
{\bm{\mathcal H}}({\bm r})&=\bm{\nabla}\times\{-y,x,0\}\mathfrak{a}(r).\label{mpot2}
\end{align}
After the substitution of these formulas into the Maxwell equations, we obtain two ordinary differential equations for $\phi(r)$ and $\mathfrak{a}(r)$,
\begin{align}
-\phi''(r)-2/r\phi'(r)=\rho(r),\label{pot1}\\
-\mathfrak{a}''(r)-4/r\mathfrak{a}'(r)=\chi(r),\label{pot2}
\end{align}
where $\rho(r)$ and $\chi(r)$ are,
\begin{align}\label{chi}
\rho(r)&=\frac{3}{4\pi a^3}\Theta(a-r)+\rho_e(r),\\
\chi(r)&=\frac{3\mu}{\pi a^4 r}\Theta(a-r)-\frac{\alpha\rho_e(r)}{r}.
\end{align}

The solutions of (\ref{pot1}) and (\ref{pot2}) can be expressed as double integrals of the source terms,
\begin{align}\label{solpot}
\phi(r)=\int_r^{\infty}\!\frac{dv}{v^2}
\int_0^v\!\!\!du\,u^2\rho(u),\\
\mathfrak{a}(r)=\int_r^{\infty}\!\frac{dv}{v^4}
\int_0^{v}\!\!\!du\,u^4\chi(u).
\end{align}

These integrals can be evaluated in closed form and the results are,
\begin{widetext}
\begin{align}\label{int}
&\phi(r)=\frac{1}{4\pi r}\left[\left(\frac{3a^2r-r^3}{2 a^3}\Theta(a-r)+\Theta(r-a)\right)-\left(1-\frac{\Gamma(1+2\gamma,2r/b)
-(2r/b)\Gamma(2\gamma,2r/b)}{\Gamma(1+2\gamma)}\right)\right],\\
&\frac{d}{dr}\phi(r)=-\frac{1}{4\pi r^2}\left[
\left(\frac{r^3}{a^3}\Theta(a-r)+\Theta(r-a)\right)
-\left(1-\frac{\Gamma(2\gamma+1,2r/b)}{\Gamma(2\gamma+1)}\right)\right],\\
&\mathfrak{a}(r)=\frac{1}{4\pi r^3}\left[\mu \left(\frac{4ar^3-3r^4}{a^4}\Theta(a-r)
+\Theta(r-a)\right)-\lambdabar\frac{\Gamma(2\gamma+2,2r/b)-(2r/b)^3\Gamma(2\gamma-1,2r/b)
-\Gamma(2\gamma-1)}
{6\Gamma(2\gamma+1)}\right],\\
&\frac{d}{dr}\mathfrak{a}(r)=-\frac{1}{4\pi r^4}\left[3\mu \left(\frac{r^4}{a^4}\Theta(a-r)+\Theta(r-a)\right)
-\lambdabar\frac{\Gamma(2\gamma+2)-\Gamma(2\gamma+2,2r/b)}
{2\Gamma(2\gamma+1)}\right],
\end{align}
where $\Gamma(z,a)$ is the incomplete gamma function \cite{nist}. The field vectors are constructed from these scalar functions according to the formulas which follow from (\ref{mpot1}) and (\ref{mpot2}),
\begin{align}
{\bm{\mathcal D}}({\bm r})=-\frac{\bm r}{r}\frac{d}{dr}\phi(r),\quad
{\bm{\mathcal H}}({\bm r})=-\frac{{\bm r}z}{r}\frac{d}{dr}\mathfrak{a}(r)+{\bm n}_z\left(r\frac{d}{dr}\mathfrak{a}(r)+2\mathfrak{a}(r)\right).
\end{align}

\begin{align}
{\bm{\mathcal D}}({\bm r})&=\frac{{\bm r}}{4\pi r^3}\left[
\left(\frac{r^3}{a^3}\Theta(a-r)+\Theta(r-a)\right)
-\left(1-\frac{\Gamma(2\gamma+1,2r/b)}
{\Gamma(2\gamma+1)}\right)\right],\label{sol1}\\
{\bm{\mathcal H}}({\bm r})&=\frac{z{\bm r}}{4\pi r^5}\left[3\mu \left(\frac{r^3}{a^3}\Theta(a-r)+\Theta(r-a)\right)
-\lambdabar\frac{\Gamma(2\gamma+2)-\Gamma(2\gamma+2,2r/b)}
{2\Gamma(2\gamma+1)}\right]\nonumber\\
-&\frac{{\bm n}_z}{4\pi r^3}\left[\mu \left(\frac{9r^4-8ar^3}{a^4}\Theta(a-r)+\Theta(r-a)\right)
-\lambdabar\frac{\Gamma(2\gamma+2)-\Gamma(2\gamma+2,2r/b)
-2(2r/b)^3\Gamma(2\gamma-1,2r/b)}
{6\Gamma(2\gamma+1)}
\right].\label{sol2}
\end{align},
\end{widetext}

\begin{figure}[b]
\begin{center}
\vspace{0.35cm}
\includegraphics[width=0.45\textwidth,
height=0.2\textheight]{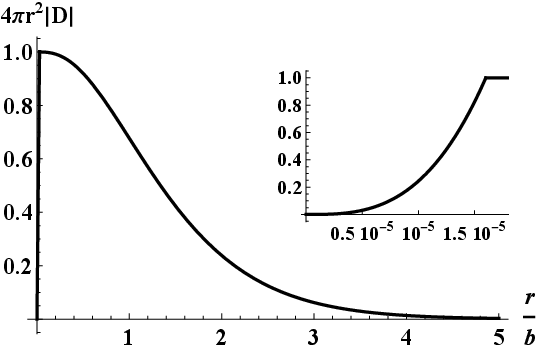}
\caption{The value of the modulus of the electric displacement field $|{\bm{\mathcal D}}|$ plotted on the atomic scale for the hydrogen atom in the ground state. In order to interpret this plot as the total amount of charge enclosed within the sphere of radius $r$ (Gauss law) we multiplied $|{\bm{\mathcal D}}|$ by $4\pi r^2$. The details of the behavior of $|{\bm{\mathcal D}}|$ close to the center are shown in the inset on the scale of the proton radius. Starting from the origin, the enclosed charge increases as $(r/a)^3$ and reaches the value of 1 at the proton radius $a$.} \label{fig1}
\end{center}
\end{figure}

\begin{figure}
\begin{center}
\vspace{0.35cm}
\includegraphics[width=0.45\textwidth,
height=0.28\textheight]{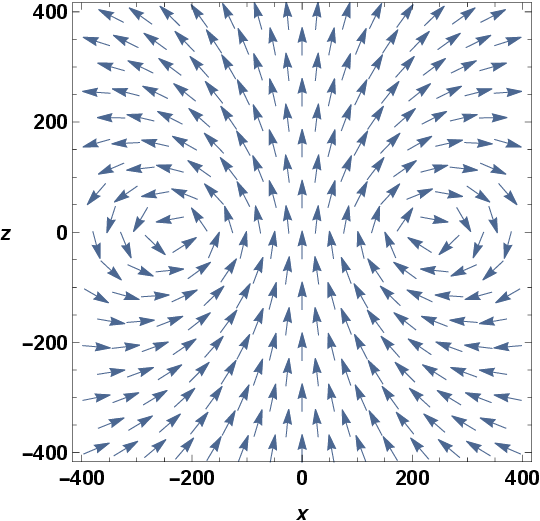}
\caption{The magnetic field surrounding the hydrogen atom in the ground state. The plot of the field configuration in the $y=0$ plane gives the complete information, owing to the rotational symmetry around the $z$ axis. The coordinates $x$ and $z$ are measured in $\lambdabar$}\label{fig2}
\end{center}
\end{figure}

Owing to the huge difference in size between the proton and the electron cloud one cannot show the complete behavior of the field vectors on a single plot. In Fig.~\ref{fig1} we show the behavior of ${\bm{\mathcal D}}$ on the atomic scale and in the inset we show the behavior of ${\bm{\mathcal D}}$ near $r=0$ on the scale of the proton radius $a$. The electric displacement field has the radial form and the magnetic field plotted in Fig.~\ref{fig2} exhibits a typical field configuration of a magnet.

In order to calculate the Zeldovich number we could use in principle the fields in the position representation and the original Zeldovich formula (\ref{avn}). However, the use of the Fourier transforms greatly simplifies the calculations. These transforms are obtained from the same purely algebraic equations (\ref{d}) and (\ref{h}) as in the classical theory. Their solutions are,
\begin{align}
{\tilde{\bm{\mathcal D}}}({\bm k})&=-i{\bm k}\frac{\tilde\rho(k)}{k^2},\label{sold}\\
{\tilde{\bm{\mathcal H}}}({\bm k})&=i\left(({\bm k}\!\cdot\!{\bm n}_z){\bm k}-k^2{\bm n}_z\right)
\frac{1}{k^3}\frac{d{\tilde\chi}(k)}{dk}.\label{solh}
\end{align}
The Fourier transform of spherically symmetric functions become one-dimensional integrals,
\begin{align}\label{ft1}
{\tilde f}(k)=\int\!\!\frac{d^3r}{(2\pi)^{3/2}}e^{-i{\bm k\cdot{\bm r}}}f(r)=\sqrt{\frac{2}{\pi}}\int_0^\infty\!\!\!\!\!dr\,r\frac{\sin(kr)}{k}f(r).
\end{align}
This integral can be evaluated in a closed form for both functions $\rho(r)$ and $\chi(r)$,
\begin{widetext}
\begin{align}
\tilde\rho({k})&=\frac{3\sin(ak)\!-\!3ak\cos(ak)}
{(2\pi)^{3/2}a^3k^3}
-\frac{\sin\left[2\gamma\arctan(bk/2)\right]}
{(2\pi)^{3/2}\gamma bk\left(1+k^2b^2/4\right)^{\gamma}},\label{sols1}\\
\tilde\chi(k)&=6\mu\frac{2ak\sin(ak)+(2-a^2k^2)\cos(ak)-2}{(2\pi)^{3/2}a^6k^4}
-\alpha\frac{2\sin\left[(2\gamma-1)\arctan(bk/2)\right]}
{(2\pi)^{3/2}\gamma(2\gamma-1)b^2k\left(1+b^2k^2/4\right)^{\gamma-1/2}},\label{sols2}\\
\frac{d\tilde\chi(k)}{dk}&=6\mu\frac{ ak(a^2k^2-8)\sin(ak)+4(a^2k^2-2)\cos(ak)+8}{(2\pi)^{3/2}a^6k^5}\nonumber\\
&-\alpha\frac{(2\gamma-1)bk\cos\left[2\gamma\arctan(bk/2)\right]
-2(2+\gamma b^2k^2)\sin\left[2\gamma\arctan(bk/2)\right]}
{(2\pi)^{3/2}2\gamma(2\gamma-1)b^2k^2/4
\left(1+b^2k^2/4\right)^{\gamma+1/2}}.\label{sols3}
\end{align}
\end{widetext}

The substitution of the expression (\ref{sold}) for ${\tilde{\bm{\mathcal D}}}$  into (\ref{nf}) gives the formula for the contribution $\mathcal{N}_Z\left[{\bm{\mathcal D}}\right]$, which is associated with the electric field, to the total value,
\begin{align}\label{phd}
\mathcal{N}_Z\left[{\bm{\mathcal D}}\right]
=2\pi\alpha\int\!\frac{d^3k}{k^3}\tilde{\rho}(k)^2.
\end{align}
The integration over $k$ cannot be done analytically and we have to resort to numerical integration. The formula for $\mathcal{N}_Z$ with $\tilde{\rho}(k)$ given by (\ref{sols1}) contains the dimensional parameters $a$ and $b$. However, this number is dimensionless so that it can only depend on a dimensionless ratio. We shall make use of this property and change the dimensional integration variable $k$ to the dimensionless variable $\kappa=bk$ and introduce the dimensionless ratio as $s=a/b=1.6\times10^{-5}$. The resulting integral is,
\begin{align}\label{elph}
&\mathcal{N}_Z\left[{\bm{\mathcal D}}\right]=\frac{\alpha}{\pi}\int_0^{\infty}\!
\frac{d\kappa}{\kappa}\\
&\!\times\!\left[\frac{3\sin(s\kappa)\!-\!3s\kappa\cos(s\kappa)}
{s^3\kappa^3}-\frac{\sin\left(2\gamma\arctan(\kappa/2)\right)}
{\gamma \kappa\left(1+\kappa^2/4\right)^{\gamma}}\right]^2\nonumber
\end{align}
and the ``electric Zeldovich number'' associated with the hydrogen atom in the ground state is $\mathcal{N}_Z\left[{\bm{\mathcal D}}\right]=0.025$. By the way, a similar number 0.02 was obtained with the use of very crude arguments in \cite{ibb2}.
The ``magnetic Zeldovich number'' for the hydrogen atom is given by the integral,
\begin{align}\label{phh}
\mathcal{N}_Z\left[{\bm{\mathcal H}}\right]
=2\pi\alpha\int\!\frac{d^3k}{k^7}(k^2-({\bm k}\!\cdot\!{\bm n}_z)^2)^2\left(\frac{d\tilde\chi({k})}{dk}\right)^2.
\end{align}
After the integration over the angles, we obtain,
\begin{align}\label{magph}
&\mathcal{N}_Z\left[{\bm{\mathcal H}}\right]=\frac{16\pi^2\alpha}{3}\int_0^{\infty}\!
\frac{dk}{k}\left(\frac{d\tilde\chi({k})}{dk}\right)^2.
\end{align}
Using  (\ref{sols2}) and after the replacement $kb=\kappa$ in (\ref{sols3}), we obtain,
\begin{widetext}
\begin{align}\label{magphe1}
\mathcal{N}_Z\left[{\bm{\mathcal H}}\right]=\frac{2\alpha}{3\pi}\!\!\int_0^{\infty}\!\!
\frac{d\kappa}{\kappa}\left[6d\frac{s\kappa(s^2\kappa^2-8)\sin(s\kappa)
+4(s^2\kappa^2-2)\cos(s\kappa)+8}{s^5\kappa^5}
-\alpha\frac{\kappa\cos\sigma-\sin\sigma}
{2\gamma(2\gamma-1)\kappa^2\left(1+\kappa^2/4\right)^\gamma}\right]^2,
\end{align}
\end{widetext}
where $d=\mu/a=0.68$ and $\sigma=2\gamma\arctan(\kappa/2)$.
Numerical integration produces a tiny number $\mathcal{N}_Z\left[{\bm{\mathcal H}}\right]=6\times10^{-5}$ which is totally negligible in comparison to the number $\mathcal{N}_Z\left[{\bm{\mathcal D}}\right]$ for the electric field. The reason for this huge difference is a rather slow motion of electrons in comparison to the speed of light. This results in the appearance of the fine structure constant in (\ref{cur}).

We also calculated the energy carried by the electromagnetic field associated with the hydrogen atom. As was to be expected the contribution of the magnetic field is negligible. The energy carried by the electric field comes almost entirely from the Coulomb field of the proton because this field is very strong at small distances. The associated energy is quite substantial; it is equal to twice the rest energy of the electron $2m_{el}c^2$. Of course, the Coulomb energy of the proton cannot be counted as a separate contribution because it is already included in the observed proton rest energy. The total field energy associated with the electronic wave function is very tiny. It is also dominated by the electric part and it is equal to $1.67\times10^{-5}m_{el}c^2$.\\

\section{The Zeldovich number for heavier atoms}

We have chosen, as an example, the atoms of noble gases because their closed shells produce spherically symmetric charge distribution which greatly simplifies the calculations. We restrict ourselves here to the calculation of the ``electric Zeldovich number'' since, as was seen in the case of the hydrogen atom, the ``magnetic Zeldovich number'' is much smaller. This number grows rapidly with the increase of the atomic number $Z$. The calculation of the exact value of the average photon number  would require the knowledge of the total wave function of mutually interacting electrons, but to obtain an order of magnitude estimate we will neglect this interaction. We also use the electron wave functions in the nonrelativistic approximation,
\begin{widetext}
\begin{align}\label{wfnr}
\psi_{nlm}(\varrho,\theta,\phi)=\sqrt{\frac{(n-l-1)!}{2n(n+l)!}}\left(\frac{2Z}{nb}\right)^{3/2}
\!\!\!\varrho^l e^{-\varrho/2}L_{n-l-1}^{2l+1}(\varrho)\,Y_l^m(\theta,\phi),
\end{align}
where $\varrho=\frac{2Zr}{nb}$. Note that we use a different font to distinguish the rescaled radial variable from the probability density.
In most textbooks on quantum mechanics the lower index of the associated Laguerre polynomial has a different meaning. Instead of $n-l-1$ it is $n+l$. We follow here the notation of {\em Mathematica} \cite{math}.
The probability density corresponding to the wave function (\ref{wfnr}) is,
\begin{align}\label{dnr}
\rho_{nlm}(\varrho,\theta,\phi)=\frac{(n-l-1)!}{2n(n+l)!}
\left(\frac{2Z}{nb}\right)^3
\!\!\!\varrho^{2l} e^{-\varrho}\left[L_{n-l-1}^{2l+1}(\varrho)\right]^2\,
Y_l^m(\theta,\phi)Y_l^m(\theta,-\phi),
\end{align}
The only dependence on the quantum number $m$ is through the spherical harmonics. Therefore, for given $n$ and $l$ we can sum up $2(2l+1)$ contributions from different values of $m$ and obtain the total contribution $\rho_{nl}$ from the fully filled $(n,l)$ shell. The additional factor of 2 is due to the two possible electron spin orientations. The density $\rho_{nl}(\varrho)$ of the electrons for the $(n,l)$ shell is,
\begin{align}\label{dnrnl}
\rho_{nl}(\varrho)=\frac{2(2l+1)}{4\pi}\frac{(n-l-1)!}{2n(n+l)!}
\left(\frac{2Z}{nb}\right)^3
\!\!\!\varrho^{2l} e^{-\varrho}\left[L_{n-l-1}^{2l+1}(\varrho)\right]^2.
\end{align}
\end{widetext}
The total charge density must also include the contribution from the nucleus $\rho_{nucl}(\varrho)$. We assume, as we have done for the proton, that the charge is distributed evenly within the sphere whose radius depends on the atomic mass number $A$.
\begin{align}\label{nuc}
\rho_{nucl}(\varrho)=\frac{3Z}{4\pi a(A)^3}\Theta(a(A)-r).
\end{align}
We shall use the commonly accepted (see, for example, \cite{brs}) formula for $a(A)$,
\begin{align}\label{aa}
a(A)=1.2\,A^{1/3}\!\!\times\!10^{-15}{\rm m}.
\end{align}
\begin{figure}[t]
\begin{center}
\vspace{0.35cm}
\includegraphics[width=0.45\textwidth,
height=0.25\textheight]{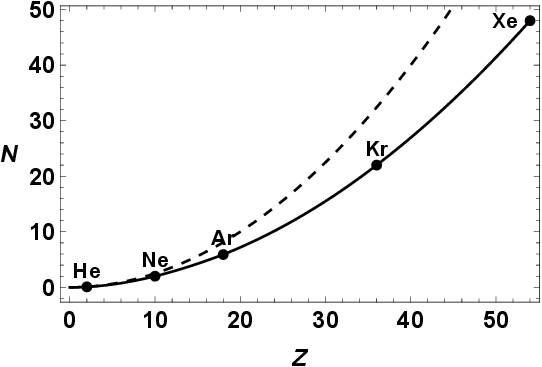}
\caption{The Zeldovich number for the atoms of noble gases.}\label{fig3}
\end{center}
\end{figure}
In order to determine the Fourier transform of the electric field ${\tilde{\bm{\mathcal D}}}({\bm k})$ we need the Fourier transform of the electron charge density for all shells and the nucleus charge density. The calculations are simple but tedious. In order to obtain the results for all stable noble gas atoms we calculate, according to the formula (\ref{ft1}), the Fourier transforms of the functions $\rho_{nl}(r)$ for all closed shells. All the integrals are evaluated analytically and they are equal to the ratios of the polynomials in $k^2$. Finally, for each atom we add up the contributions from the relevant shells and evaluate the final value (\ref{phd}) of the Zeldovich number.

The results are shown in Fig.~\ref{fig3}. As was to be expected the dependence on $Z$ is almost quadratic (dashed line). The deviation is due to the increase of the nucleus radius with the increasing atomic mass number $A$. This effect diminishes the strength of the electric field at the center of the atom.

\section{Conclusions}

The results presented in this work have probably no immediate applications. They offer, however, a fresh point of view by introducing a new universal dimensionless measure of the overall strength of the electromagnetic field: the same in classical and in quantum domain. For a free quantized electromagnetic field, this measure gives the number of photons. The Zeldovich number can also be viewed as a measure of the strength of the sources of the electromagnetic field. There is one property of $\mathcal{N}_Z$ which is worth stressing: it is an intensive and not an extensive property of the system. It does not depend on the size of the system but only on the {\em dimensionless ratios} of various parameters to the overall size.


\begin{thebibliography}{00}
\bibitem{zeld} Ya. B. Zeldovich, Number of quanta as an invariant of the classical electromagnetic field, Dokl. Acad. Sci. USSR, {\bf 163}, 1359 (1965), (In Russian).
\bibitem{pcp0} R. Passante, G. Compagno and F. Persico, Cloud of virtual photons in the ground state of the hydrogen atom, Phys. Rev. A {\bf 31}, 2827 (1985).
\bibitem{pcp} F. Persico, G. Compagno and R. Passante, Ground State Photons Dressing Atoms and Molecules, {\em Quantum Optics IV} edited by J. D. Harvey and D. F. Walls, (Springer, Berlin, 1986), p. 172.
\bibitem{cpps} G. Compagno, R. Passante, F. Persico, and G. M. Salamone, Cloud of virtual photons surrounding a nonrelativistic electron, Acta Phys. Polon. {\bf 85}, 667 (1994).
\bibitem{cppp} G. Compagno, G. M. Palma, R. Passante and F. Persico, Atoms dressed and partially dressed by the zero-point fluctuations of the electromagnetic field, J. Phys. B {\bf 8}, 1105 (1995).
\bibitem{tr} T. Radozycki, The electromagnetic virtual cloud of the ground-state hydrogen atom-a quantum field theory approach, J. Phys. A {\bf 23}, 4911 (1990).
\bibitem{gross} L. Gross, Norm invariance of mass-zero equations under the conformal group, J. Math. Phys. {\bf 5}, 687 (1964).
\bibitem{rmp} L.C.B. Crispino, A. Higuchi and G.E.A. Matsas, The Unruh effect and its applications, Rev. Mod. Phys. {\bf 80}, 787 (2008).
\bibitem{pwf} I. Bialynicki-Birula, Photon wave function, {\em Progress in Optics}, edited by E. Wolf, Elsevier, Amsterdam, {\bf 36}, 1 (1996); (see also ArXiv: quant-ph/0508202).
\bibitem{bb1} I. Bialynicki-Birula and Z. Bialynicka-Birula, The role of the Riemann-Silberstein vector in classical and quantum theories of electromagnetism, J. Phys. A: Math. Theor. {\bf 46} 053001 (2013).
\bibitem{bb} I. Bialynicki-Birula and Z. Bialynicka-Birula, Three measures of fidelity for photon states, Phys. Rev. A {\bf 102}, 042201 (2020).
\bibitem{ibb1} I. Bialynicki-Birula, The Wigner functional of the electromagnetic field, Opt. Comm. {\bf 179}, 237 (2000).
     There are some misplaced factors of 2 in this reference that were corrected in \cite{ibb2}.
\bibitem{ibb2} I. Bialynicki-Birula, The structure of the vacuum and the photon number, in {\em Decoherence and Entropy in Complex Systems}, Lect. Notes Phys., {\bf 633}, 287, (Springer, Berlin, 2004).
\bibitem{ibb3} I. Bialynicki-Birula, Relativistic Wigner functions, EPJ Web of Conf. {\bf 78}, 01001 (2014).
\bibitem{jack} J. D. Jackson, {\em Classical Electrodynamics}, (John Wiley \& Sons, Hoboken, NJ, 1999), p.181.
\bibitem{bd} J. D. Bjorken and S. D. Drell, {\em Relativistic Quantum Mechanics}, (McGraw-Hill, New York, 1965), p.55.
\bibitem{march} L. Marchildon, Quantum Mechanics, (Springer, Berlin, 2002) p. 211.
\bibitem{nist} NIST Handbook of Mathematical Functions (University Press, Cambridge, 2010) p. 174.
\bibitem{math} Wolfram Research, Inc., Mathematica, Version 13.1, Champaign, (2022).
\bibitem{brs} J. L. Basdevant, J. Rich, and M. Spiro, {\em Fundamentals in Nuclear Physics}, (Springer, Berlin, 2005) p. 151.
\end{thebibliography}
\end{document}